\documentclass[epj]{webofc}
\usepackage[utf8]{inputenc}
\usepackage[varg]{txfonts}   
\usepackage{booktabs}
\usepackage{xcolor}
\definecolor{darkred}{rgb}{0.4,0.0,0.0}
\definecolor{darkgreen}{rgb}{0.0,0.4,0.0}
\definecolor{darkblue}{rgb}{0.0,0.0,0.4}
\usepackage[bookmarks,linktocpage,colorlinks,
    linkcolor = darkred,
    urlcolor  = darkblue,
    citecolor = darkgreen]{hyperref}
\usepackage{subfigure}
\wocname{EPJ Web of Conferences}
\woctitle{Lattice2017}
\graphicspath{{figs/}}
\DeclareMathOperator{\diag}{diag}
\DeclareMathOperator{\sign}{sign}
\DeclareMathOperator{\tr}{tr}
\newcommand{\hm}{u}
\newcommand{\hx}{x} 
\newcommand\zm{\text{zm}}
\newcommand\nzm{\text{nzm}}
\renewcommand\epsilon\varepsilon
\renewcommand\phi\varphi
\begin{document}
%
\selectlanguage{english}
\title{%
  \boldmath Chiral condensate and Dirac spectrum of one- and
  two-flavor QCD at nonzero $\theta$-angle
}
\author{%
\firstname{Mario} \lastname{Kieburg}\inst{1} \and
\firstname{Jacobus} \lastname{Verbaarschot}\inst{2}\fnsep\thanks{Speaker, \email{jacobus.verbaarschot@stonybrook.edu}} \and
\firstname{Tilo} \lastname{Wettig}\inst{3} 
}
\institute{%
  Department of Physics, University of Bielefeld, 33501 Bielefeld, Germany
  \and
  Department of Physics and Astronomy,  Stony Brook University, New York 11794, USA
  \and
  Department of Physics, University of Regensburg, 93040 Regensburg, Germany
}
\abstract{%
  In previous work we showed that the chiral condensate of one-flavor
  QCD exhibits a Silver Blaze phenomenon when the quark mass crosses
  $m=0$: the chiral condensate remains constant while the quark mass
  crosses the spectrum of the Dirac operator, which is dense on the
  imaginary axis. This behavior can be explained in terms of
  exponentially large cancellations between contributions from the
  zero modes and from the nonzero modes when the quark mass is
  negative.  In these proceedings we show that a similar Silver Blaze
  phenomenon takes places for QCD with one flavor and arbitrary
  $\theta$-angle, and for QCD with two flavors with different quark
  masses $m_1$ and $m_2$.  In the latter case the chiral condensate
  remains constant when $m_1$ crosses zero at fixed $m_2>0$ until the
  Dashen point $m_1 = -m_2$ is reached, where the chiral condensate
  has a discontinuity. In terms of contributions from the Dirac
  spectrum the shift of the discontinuity from $m_1=0$ to $m_1=-m_2$
  also arises from exponentially large cancellations between the zero
  and nonzero modes when $m_1m_2 <0$. All calculations are performed
  in the microscopic or $\epsilon$-domain of QCD. Results for
  arbitrary $\theta$-angle are discussed as well.%
}
\maketitle
\section{Introduction}\label{intro}

Although the experimental value of the $\theta$-angle is consistent
with zero, the $\theta$-dependence of the QCD partition function has
attracted a great deal of attention since the early days of QCD
\cite{tHooft:1977rfq,Gaiotto:2017yup} and also in lattice QCD
\cite{Azcoiti:2002vk,Azcoiti:2017mxl}. One reason is the close
connection with topology. For example, the topological susceptibility
can be obtained from the second derivative of the partition function
with respect to $\theta$. The $\theta$-dependence of the QCD partition
function also makes it possible to isolate the partition function at
fixed topological charge \cite{Leutwyler:1992yt}, which was the
starting point for the connection between chiral random matrix theory
and QCD \cite{Shuryak:1992pi}.

\begin{figure}[t!]
  \centerline{\includegraphics{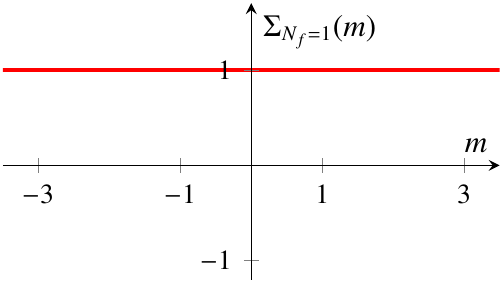}\hspace*{.1\linewidth}\includegraphics{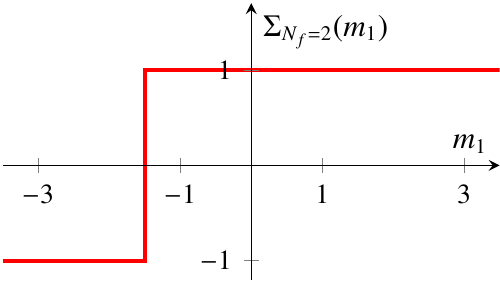}}
  \smallskip
  \caption{Behavior of the chiral condensate (in suitable units) for
    $N_f=1$ (left) and $N_f=2$ (right, with $m_2=1.5$).}
  \label{fig1}
\end{figure}

\begin{figure}[b!]
  \centerline{\includegraphics{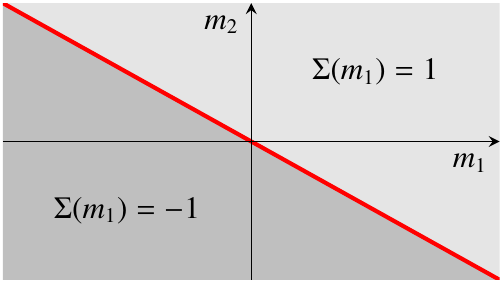}\hspace*{.1\linewidth}
    \includegraphics{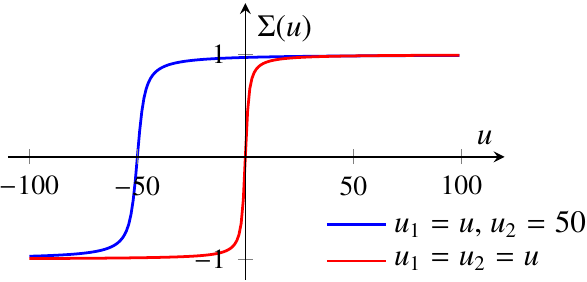}}
  \caption{Left: phase diagram of the chiral condensate for QCD with
    two flavors in the plane of the two quark masses at $\theta
    =0$. Right: mass dependence of the chiral condensate at $\theta=0$
    when both masses are equal (red curve) and when one of the masses
    is kept fixed (blue curve).  We used the notation
    $\hm = mV\Sigma$.}
\label{fig2}
\end{figure}

Another reason for considering a nonzero $\theta$-angle is to study
the effect of negative quark masses, which is equivalent to
$\theta=\pi$. For one flavor the chiral condensate is constant when
the quark mass changes sign (see Fig.~\ref{fig1}).  This raises the
question of what happens to the spectrum of the Dirac operator at
$\theta=\pi$ or when the quark mass crosses the imaginary axis
\cite{Creutz:2006ts} and how one can reconcile this property with the
Banks-Casher relation \cite{Banks:1979yr}.  For two flavors, the
Dashen phenomenon
\cite{Dashen:1970et,Akemann:2001ir,Lenaghan:2001ur,Aoki:2014moa,Horkel:2015oa}
has a long history. It occurs when the sum of the quark masses
vanishes, or for equal quark masses at $\theta=\pi$.  At this point,
the chiral condensate changes sign, and one of the questions we have
been interested in is to explain the phase diagram shown in the plane
of the two quark masses (see Fig.~\ref{fig2}) in terms of the spectrum
of the Dirac operator.

In the $\epsilon$-domain of QCD the exact analytical form of the
spectral density of the Dirac operator as well as the partition
function at fixed topological charge $\nu$ are given by expressions in
terms of Bessel functions that can be derived from chiral random
matrix theory
\cite{Verbaarschot:1994qf,Verbaarschot:2000dy,Verbaarschot:2009jz}. It
is possible to evaluate the sums over $\nu$
\cite{Verbaarschot:2014upa,Kieburg:2017}, which provides us with exact
analytical results for the spectral density at fixed $\theta$-angle
and allows us to study the connection between the spectral density and
the chiral condensate at fixed $\theta$-angle. In
\cite{Verbaarschot:2014upa,Verbaarschot:2014qka} this program was
carried out for $N_f =1$ and $\theta$ equal to 0 or $\pi$. In these
proceedings we will investigate the same questions for both $N_f =1$ and $N_f =2$
at arbitrary $\theta$-angle. More details will be discussed in a
forthcoming paper \cite{Kieburg:2017}.

\section{Chiral condensate for one- and two-flavor QCD}

The $\theta$-dependence of the one-flavor QCD partition function is given by
\begin{align}
  Z(m,\theta) = \sum_\nu e^{i\nu\theta}Z_\nu(m)\quad\text{with}\quad
  Z_\nu(m)=\biggl\langle \prod_{k}(i\lambda_k +m) \biggr\rangle_\nu
  = m^{|\nu|} \biggl\langle \prod_{\lambda_k>0}(\lambda_k^2+m^2) \biggr\rangle_\nu\,,
\end{align}
where the average is over gauge field configurations with topological
charge $\nu$, $m$ is the quark mass (which we assume to be real), and
the $i\lambda_k$ are the eigenvalues of the anti-Hermitian Dirac
operator (which are either zero or occur in pairs $\pm
i\lambda_k$). Note that a negative quark mass can be interpreted as
$\theta\to\theta+\pi$ and visa versa.  The partition function for
fixed topological charge follows by Fourier inversion, 
\begin{align}
  Z_\nu(m) = \frac1{2\pi} \int_{-\pi}^\pi d\theta\, e^{-i\nu\theta} Z(m,\theta)\,.
\end{align}
For one-flavor QCD, chiral symmetry is broken by the anomaly, there is
no spontaneous symmetry breaking, and there are no Goldstone bosons.
The mass dependence of the one-flavor QCD partition function is given
by \cite{Leutwyler:1992yt}
\begin{align}
  \label{eq:Z1}
  Z(m,\theta) = e^{mV\Sigma \cos\theta + O(m^2 V)}\,,
\end{align}
where $V$ is the volume and $\Sigma$ is the absolute value of the
chiral condensate in the limit $m=0$ and $\theta=0$.  The chiral
condensate is defined as
\begin{align}
  \Sigma(m) = -\langle \bar qq\rangle
  = \frac 1V \frac d{dm} \log Z(m)
  = \left \langle \frac 1V \sum_k \frac 1{i\lambda_k+m} \right \rangle,
  \label{cond}
\end{align}
where the average includes the fermion determinant.  This definition
is valid for both fixed $\nu$ and fixed $\theta$.  Equations
\eqref{eq:Z1} and \eqref{cond} imply that the condensate as a function
of $\theta$ vanishes for $ \theta = \pi/2$, which will be discussed in
detail in the next section.

The two-flavor partition function is
\begin{align}  
  Z(m_1,m_2,\theta)= \sum_\nu e^{i\nu \theta} (m_1m_2)^{|\nu|}
  \biggl\langle \prod_{\lambda_k\ne0}(i\lambda_k+m_1)(i\lambda_k+m_2)\biggr\rangle_\nu\,,
\end{align}
where we again assume that the masses are real.  The behavior of the
chiral condensate shown in Fig.~\ref{fig2} can be understood simply in
terms of the chiral Lagrangian. The mean-field result for the
two-flavor partition function at $\theta=0$ is given by
\begin{align}  
  Z(m_1,m_2,\theta=0)=\int_{U\in\text{SU}(2)} dU\,e^{V\Sigma\tr[\diag(m_1,m_2)(U+U^{-1})]}\,. 
\end{align}
In the thermodynamic limit $U$ aligns itself with the mass term,
resulting in \cite{Leutwyler:1992yt}
\begin{align}
  Z(m_1,m_2,\theta=0) = e^{V\Sigma|m_1 +m_2|}\,.
\end{align}
The chiral condensate is then
\begin{align}
  \Sigma(m_1,\theta=0) = -\langle \bar q_1q_1\rangle 
  = \frac 1V \frac d{dm_1} \log Z(m_1,m_2,\theta=0)
  = \Sigma\sign(m_1+m_2)\,, 
\end{align}
which is the behavior shown in Figs.~\ref{fig1} (right) and
\ref{fig2}.

\begin{figure}[t!]
  \begin{minipage}[t]{.47\linewidth}
    \centerline{\includegraphics{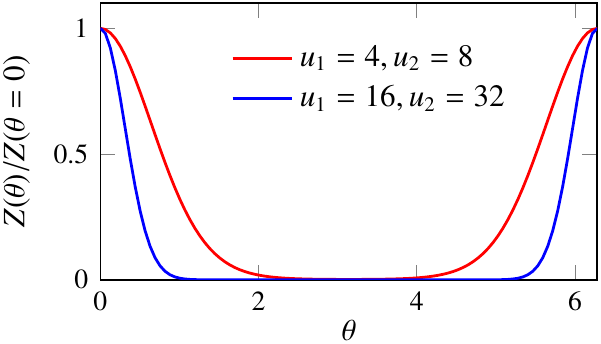}}
    \caption{Severity of the sign problem for two flavors as given by
      ${Z(m_1,m_2,\theta)}/{Z(m_1,m_2,\theta= 0)}$, to leading order
      in chiral perturbation theory.}
    \label{fig3}
  \end{minipage}\hfill
  \begin{minipage}[t]{.47\linewidth}
    \centerline{\raisebox{7pt}{\includegraphics{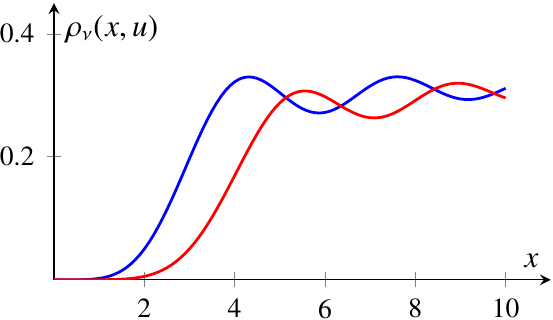}}}
    \caption{One-flavor microscopic spectral density for $\nu=2$ and
      $\hm=mV\Sigma=1$ (red) compared to the quenched result for $\nu=2$
      (blue).}
    \label{fig4}
  \end{minipage}
\end{figure}

Both for one-flavor QCD and for two-flavor QCD with different masses,
the chiral condensate does not have a discontinuity when the quark
mass crosses the line of eigenvalues, which also for nonzero $\theta$
are dense on the imaginary axis. This remarkable property seems to
violate the Banks-Casher relation, which is derived starting from the
RHS of \eqref{cond}. Due to the symmetries of the $\lambda_k$, the sum
changes sign when $m\to -m$.  Apparently, this sign change gets
compensated by the sign of the fermion determinant in the measure,
which can be negative when $m <0$ or $m_1m_2 < 0$.  A similar ``Silver
Blaze phenomenon'' arises for QCD at nonzero chemical potential, where
the chiral condensate remains constant while the Dirac spectrum is
strongly altered by the chemical potential
\cite{Cohen:2003kd,Osborn:2005ss,Kanazawa:2011tt}.  The motivation to
study QCD at nonzero $ \theta $-angle came from the desire to
understand this Silver Blaze behavior of the chiral condensate in
terms of the Dirac spectrum, but also to clarify statements in the
literature on the spectral density of the Dirac operator
\cite{Creutz:2006ts}. Since QCD at nonzero $\theta$-angle also has a
severe sign problem, see Fig.~\ref{fig3} (left), we expect that the
explanation of this behavior will be similar to that of QCD at nonzero
chemical potential, where it was found that the apparent discontinuity
of the chiral condensate can be moved by a strongly oscillating and
exponentially increasing spectral density \cite{Osborn:2005ss}.

\section{\boldmath Dirac spectrum and chiral condensate for $N_f=1$}
  
The spectral density of the Dirac operator at fixed $\theta$-angle is given by
\begin{align}
  \rho(\lambda,m,\theta)
  = \sum_\nu P_\nu \rho_\nu(\lambda,m)\,, \qquad
  P_\nu = \frac{e^{i\nu\theta } Z_\nu(m)}{Z(m,\theta)}\,,
  \label{rhonu}
\end{align}
where $P_\nu$ is the statistical weight to find a gauge field
configuration with topological charge $\nu$.
Note that for $\theta\ne0$ and/or $m<0$ the partition function is not
positive definite and $ \rho(\lambda,m,\theta)$ may become negative.  The
chiral condensate is still given by
\begin{align}
  \Sigma(m)=\frac1V\int_{-\infty}^\infty d\lambda\, \frac{\rho(\lambda)}{i\lambda+m}\,.
\end{align}
The spectral density $\rho(\lambda)$ will be decomposed into a
zero-mode part and a nonzero-mode part, the latter being the sum of a
``quenched'' and a ``dynamical'' part, 
\begin{align}
  \rho(\lambda)
  &= \rho_\zm(\lambda) + \rho_\nzm(\lambda)\notag\\
  &= \delta(\lambda) \sum_\nu |\nu|P_\nu + \rho_q(\lambda) + \rho_d(\lambda)\,.
\end{align}
The quenched part at fixed $\theta$ is obtained via \eqref{rhonu} from
the quenched part at fixed $\nu$. While the latter does not depend on
the mass, $\rho_q(\lambda)$ at fixed $\theta$ does since $P_\nu$ does,
see \eqref{eq:rhoq} below.

We will do our calculations in the microscopic domain of QCD, where
analytical results for the spectral density are given by chiral random
matrix theory.  In this domain, also known as the $\epsilon$-domain,
the quark mass and the Dirac eigenvalues scale as $m \sim 1/V$ and
$\lambda \sim 1/V$ in the thermodynamic limit.  Correction terms will
enter when $ m, \lambda$ are no longer small compared to
$1/\Lambda_{\rm QCD}\sqrt V$.  The one-flavor spectral density in the
$\epsilon$-domain is given by \cite{Damgaard:1997ye,Wilke:1997gf}
\begin{align}
  \rho_\nu(\hx,\hm) = |\nu|\delta(\hx)
  + \frac{|\hx|} 2[J_{\nu}^2(\hx) -J_{\nu+1}(\hx) J_{\nu-1}(\hx)]
  - \frac{|\hx|}{\hx^2+\hm^2}\left[\hx J_\nu(\hx) J_{\nu+1}(\hx)
  + \hm \frac{I_{\nu+1}(\hm)}{I_\nu(\hm)} J_\nu^2(\hx) \right],
\end{align}
where $ \hx \equiv \lambda \Sigma V$ and $ \hm \equiv m \Sigma V$.
An example is shown in Fig.~\ref{fig4}.
\begin{figure}[t!]
  \centerline{\includegraphics[height=40mm]{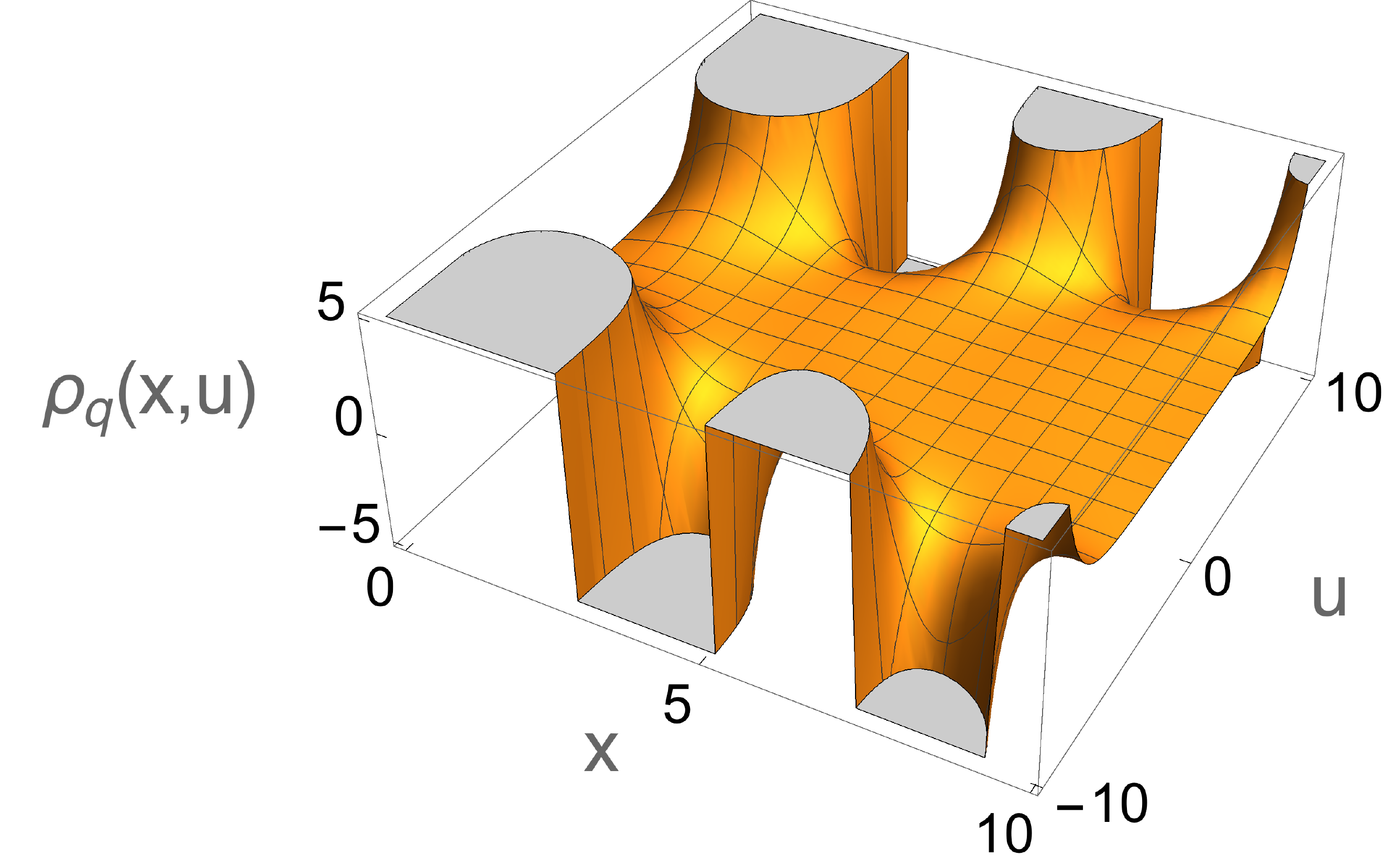}\hspace*{10mm}
    \includegraphics[height=40mm]{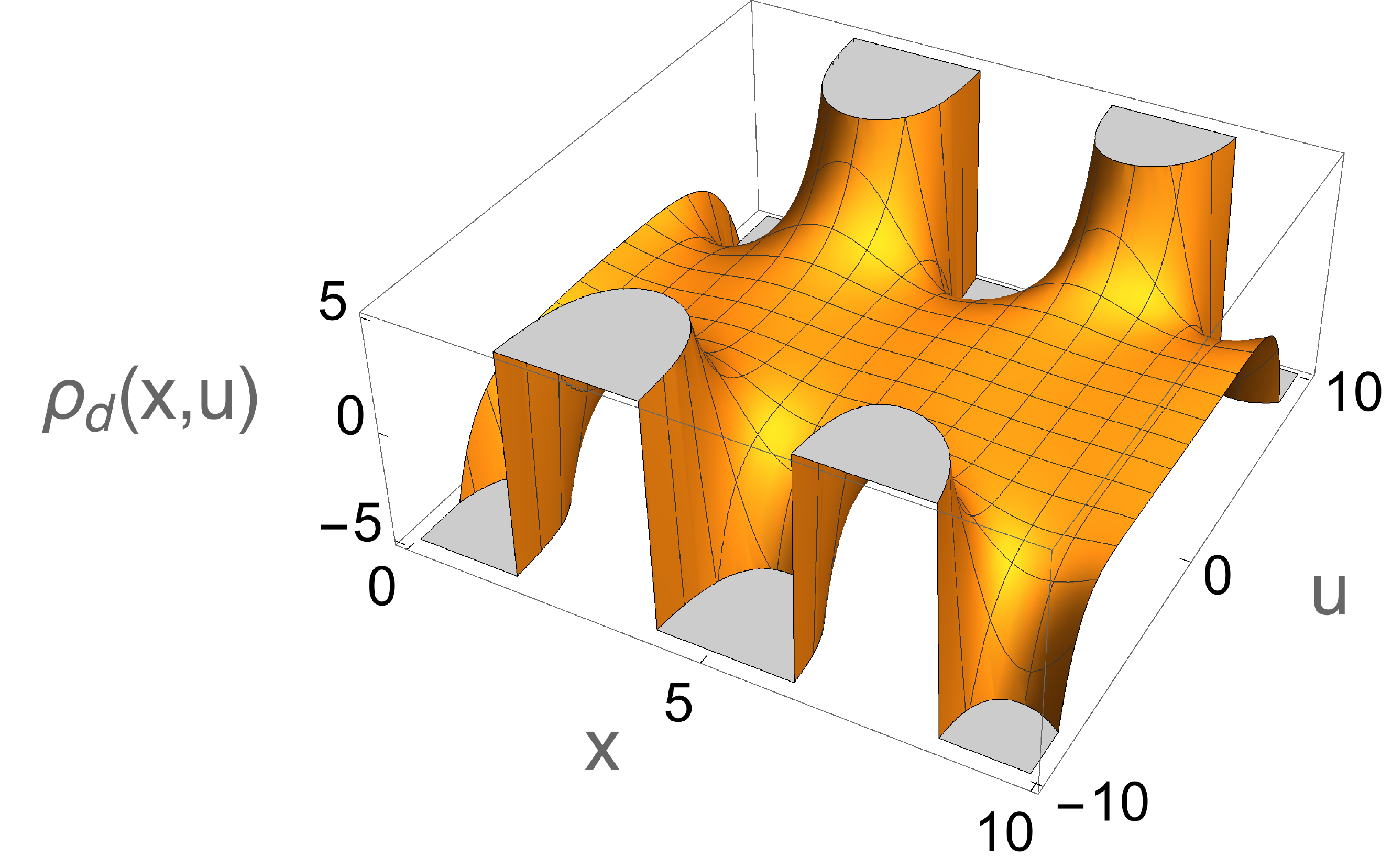}}
  \caption{Spectral density of the Dirac operator for one-flavor QCD
    at $\theta = \pi/2$. Left: quenched part, right: dynamical part.}
  \label{fig5}
\end{figure}
The spectral density at fixed $\theta$-angle is obtained by summing
over $\nu$, resulting in \cite{Kieburg:2017}
\begin{align}
  \rho_\zm(\hx,\hm,\theta) 
  &= -\delta(\hx) \int_{-\pi}^\pi \frac {d\phi}{4\pi\sin^2\frac\phi2}\,
    \frac {e^{\hm\cos(\theta+\phi)}-e^{\hm\cos\theta}}
    { e^{\hm\cos\theta}}\,, \label{eq:rhozm}\\
  \rho_q(\hx,\hm,\theta) 
  &= \int_{-\pi}^\pi\frac{d\phi}{4 \pi\sin \frac \phi 2}\,
    \frac{e^{\hm\cos(\theta+\phi)}}{e^{\hm\cos\theta}} 
    J_1\bigl(2|\hx| \sin \tfrac \phi 2\bigr)\,, \label{eq:rhoq}\\
  \rho_d(\hx,\hm,\theta) 
  &= \frac {|\hx|}{\hx^2+\hm^2} \int_{-\pi}^\pi\frac{d\phi}{2 \pi}\,
    \frac{e^{\hm\cos(\theta+\phi)}}{e^{\hm\cos\theta} }
    \left[i\hx e^{i\phi/2} J_1\bigl(2\hx \sin \tfrac \phi 2\bigr)
    -\hm e^{-i(\theta+\phi)}J_0\bigl(2\hx \sin\tfrac  \phi 2\bigr)\right],
\end{align}
where \eqref{eq:rhozm} is a principal-value integral.  In
Fig.~\ref{fig5} we show $\rho_q$ and $\rho_d$ for $\theta = \pi/2$ as
a function of $\hm$ and $\hx$. For increasing absolute value of $\hm$,
we observe oscillations with an amplitude that grows exponentially
with the volume and a period on the order of the inverse volume.
\begin{figure}[b!]
  \centerline{\includegraphics{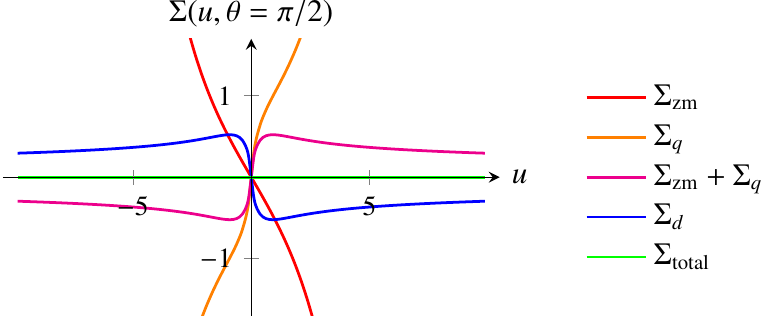}}
  \caption{Contributions to the chiral condensate for one-flavor QCD
    at $\theta=\pi/2$.}
  \label{fig6}
\end{figure}
The chiral condensate corresponding to this spectral density should
vanish (since $\theta=\pi/2$)! In Fig.~\ref{fig6} we see that both the
quenched and the zero-mode contributions to the chiral condensate
increase exponentially with the volume, while the dynamical part
remains finite. The leading-order contributions of an asymptotic
expansion in $1/V$ cancel \cite{Kanazawa:2011tt},
\begin{align}
  \Sigma_q \sim \frac{\sign(m)e^{|m|V\Sigma}}{\sqrt{2\pi}|m|^{3/2}}\bigl[1+O(1/mV\Sigma)\bigr]\,,
  \qquad
  \Sigma_\zm \sim -\frac{\sign(m)e^{|m|V\Sigma}}{\sqrt{2\pi}|m|^{3/2}}\bigl[1+O(1/mV\Sigma)\bigr]\,.
\end{align}
In fact this cancellation holds to all orders in $1/mV\Sigma$, and the
sum of the quenched part and the zero-mode part is finite
\cite{Kieburg:2017}.

\begin{figure}[t!]
  \centerline{\includegraphics[height=40mm]{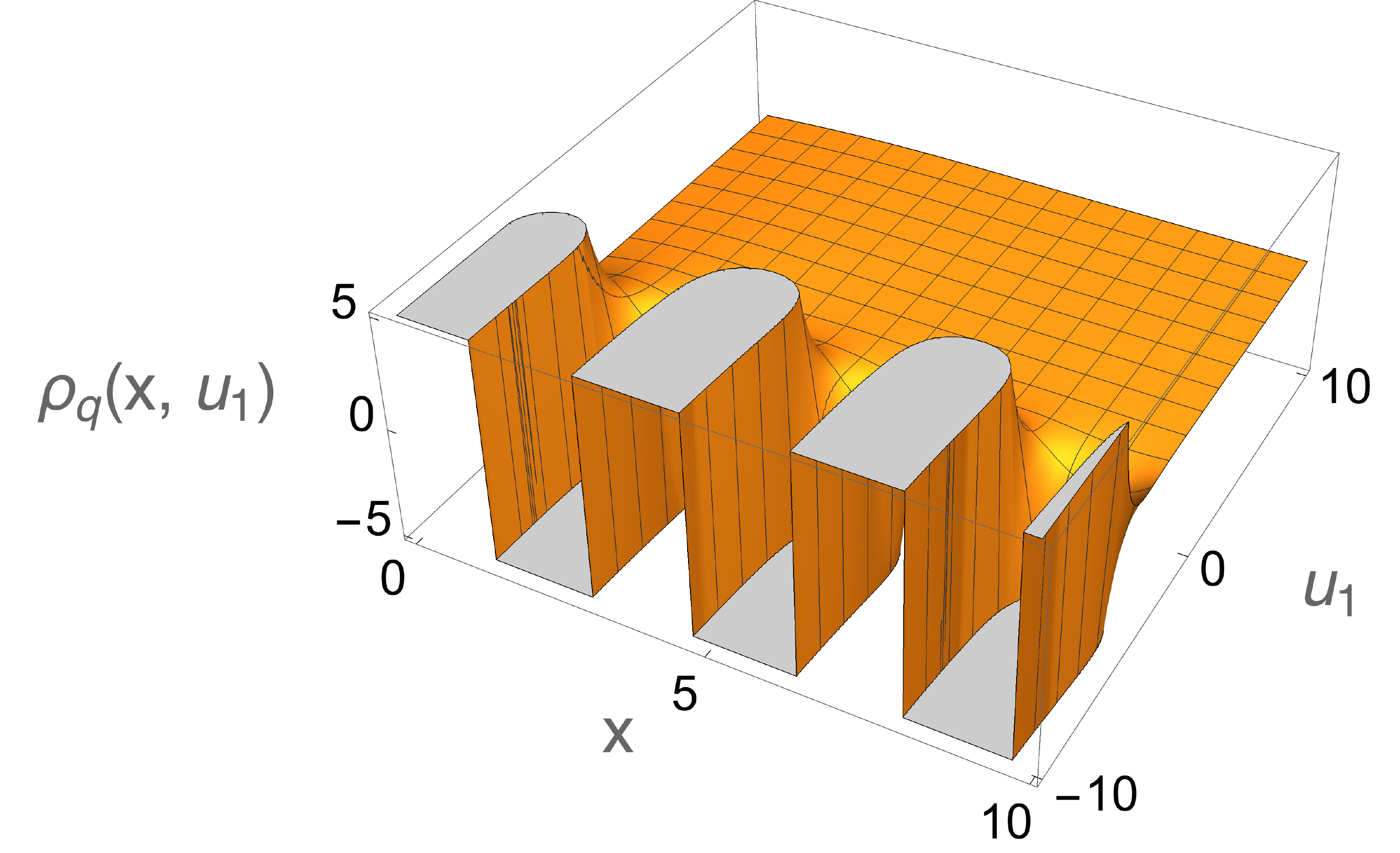}\hspace*{10mm}
    \includegraphics[height=40mm]{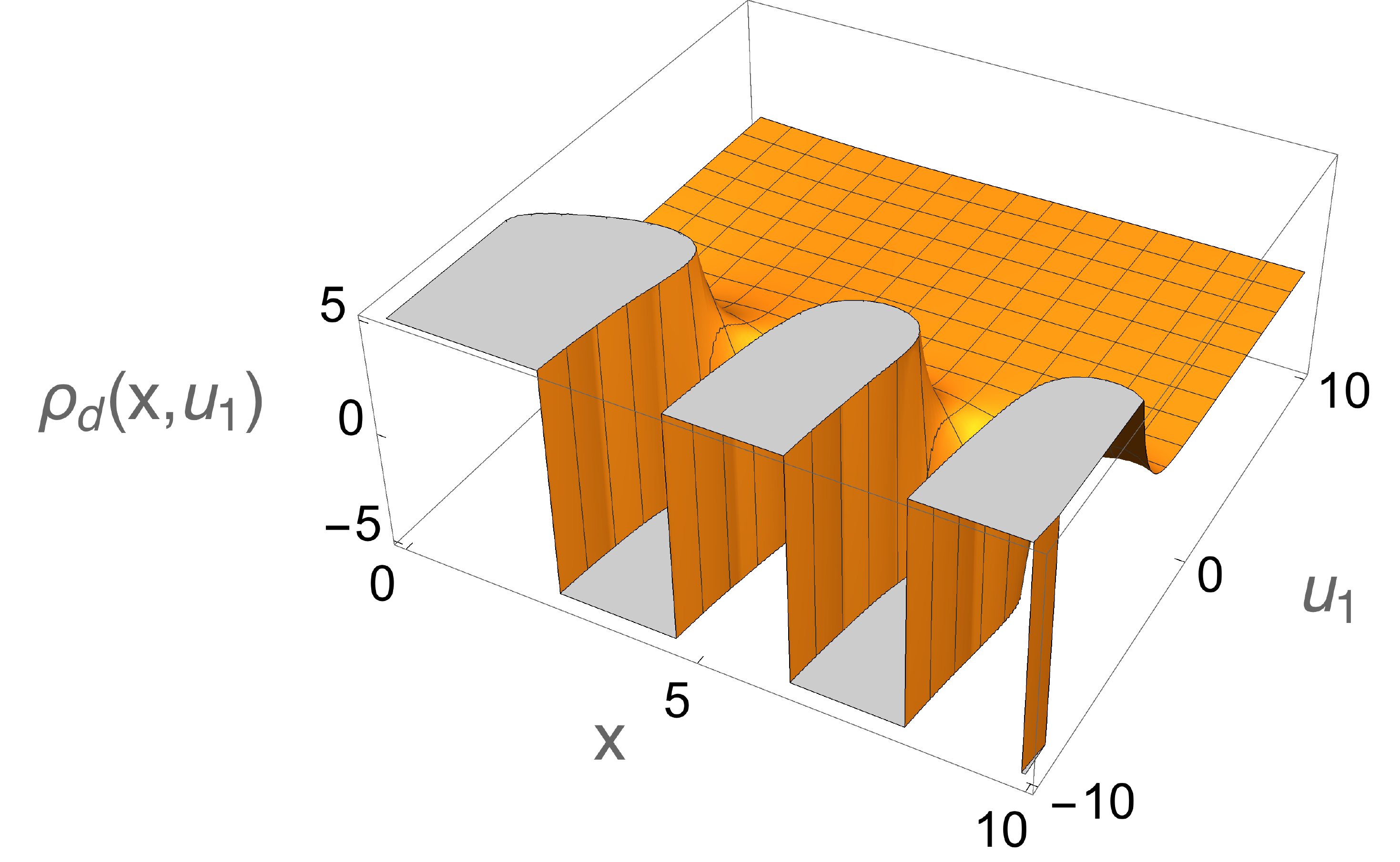}}
  \caption{Spectral density of the QCD Dirac operator for two flavors
    and $\theta=0$ with $\hm_2=10$. Left: quenched contribution,
    right: correction induced by the fermion determinant.}
\label{fig7}
\end{figure}

\section{\boldmath Dirac spectrum and chiral condensate for $ N_f =2$}

\begin{figure}
  \begin{center}
    \includegraphics[width=57mm]{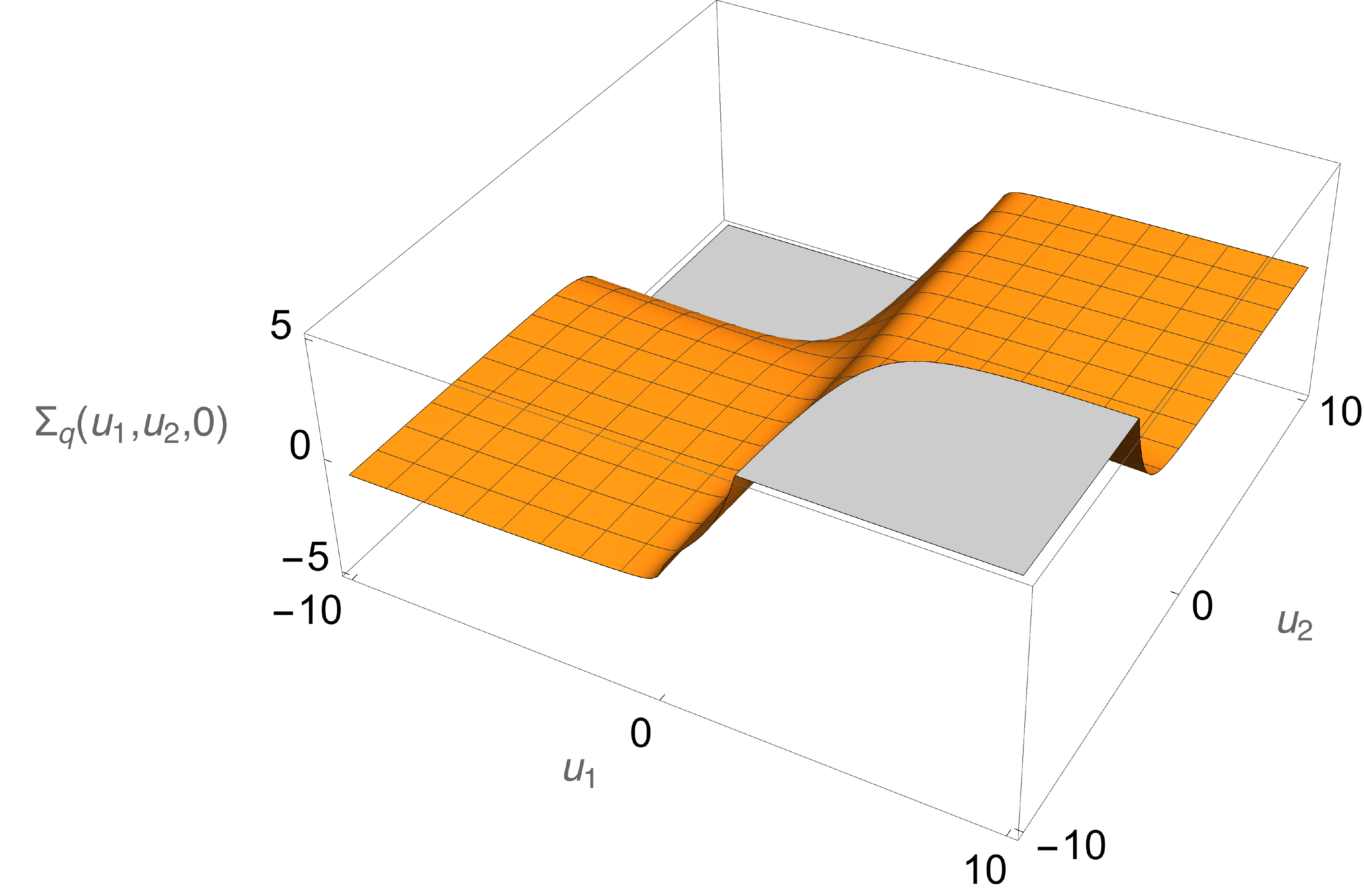}\hspace*{10mm}
    \includegraphics[width=57mm]{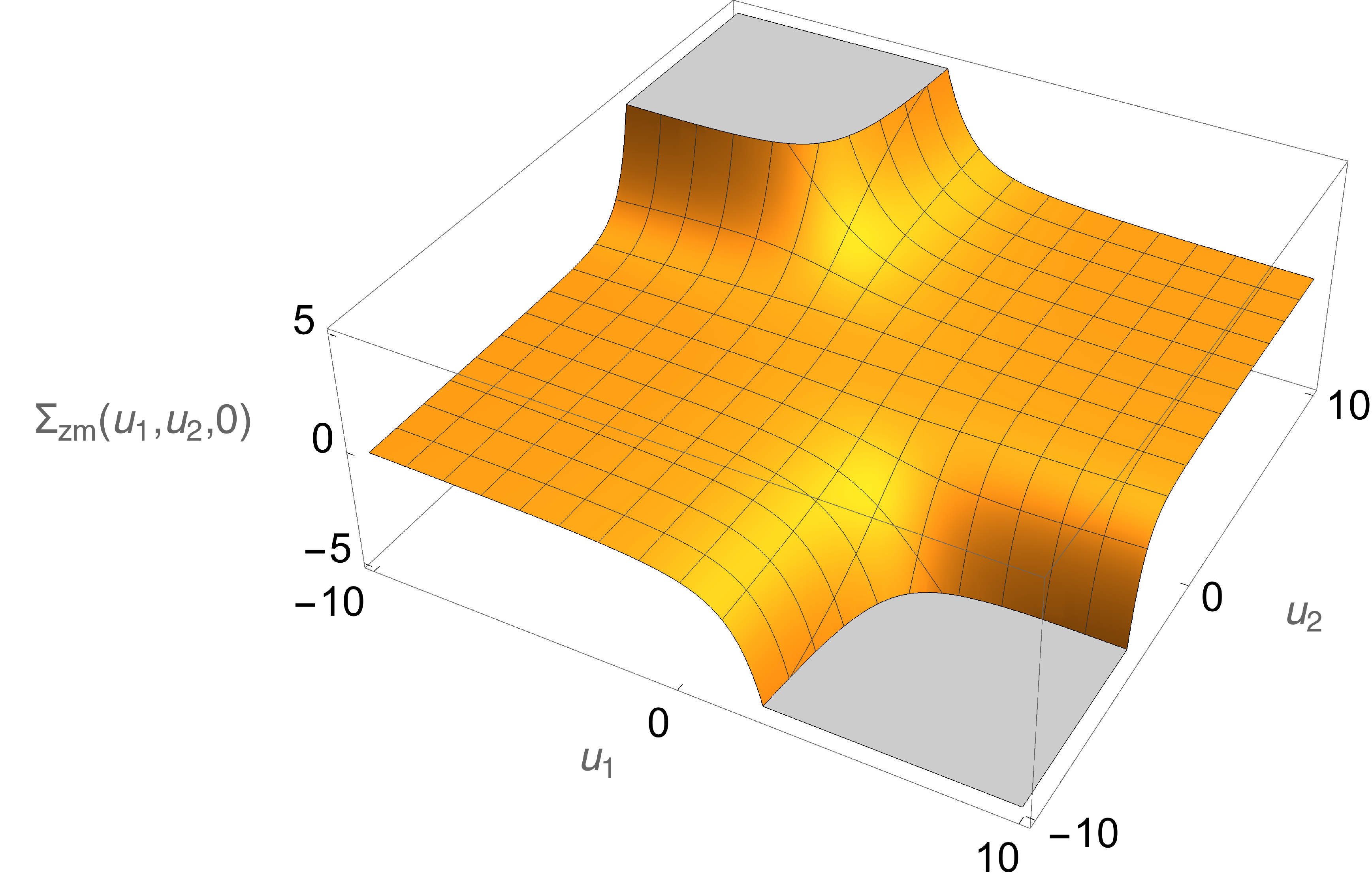}\\[-1mm]
    \includegraphics[width=57mm,angle=0]{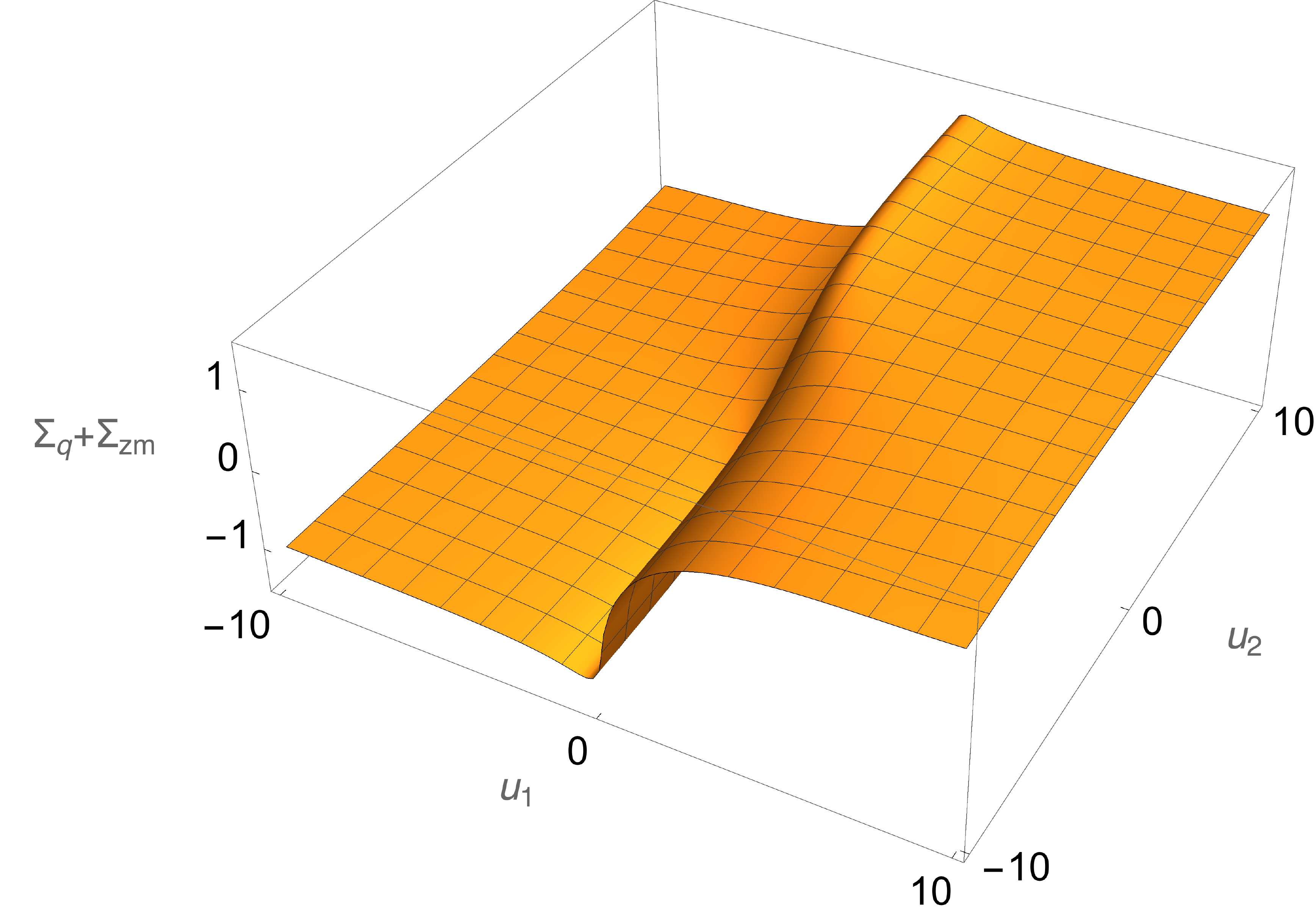}\\[-3mm]
    \hspace*{-5mm}\includegraphics[width=57mm,angle=0]{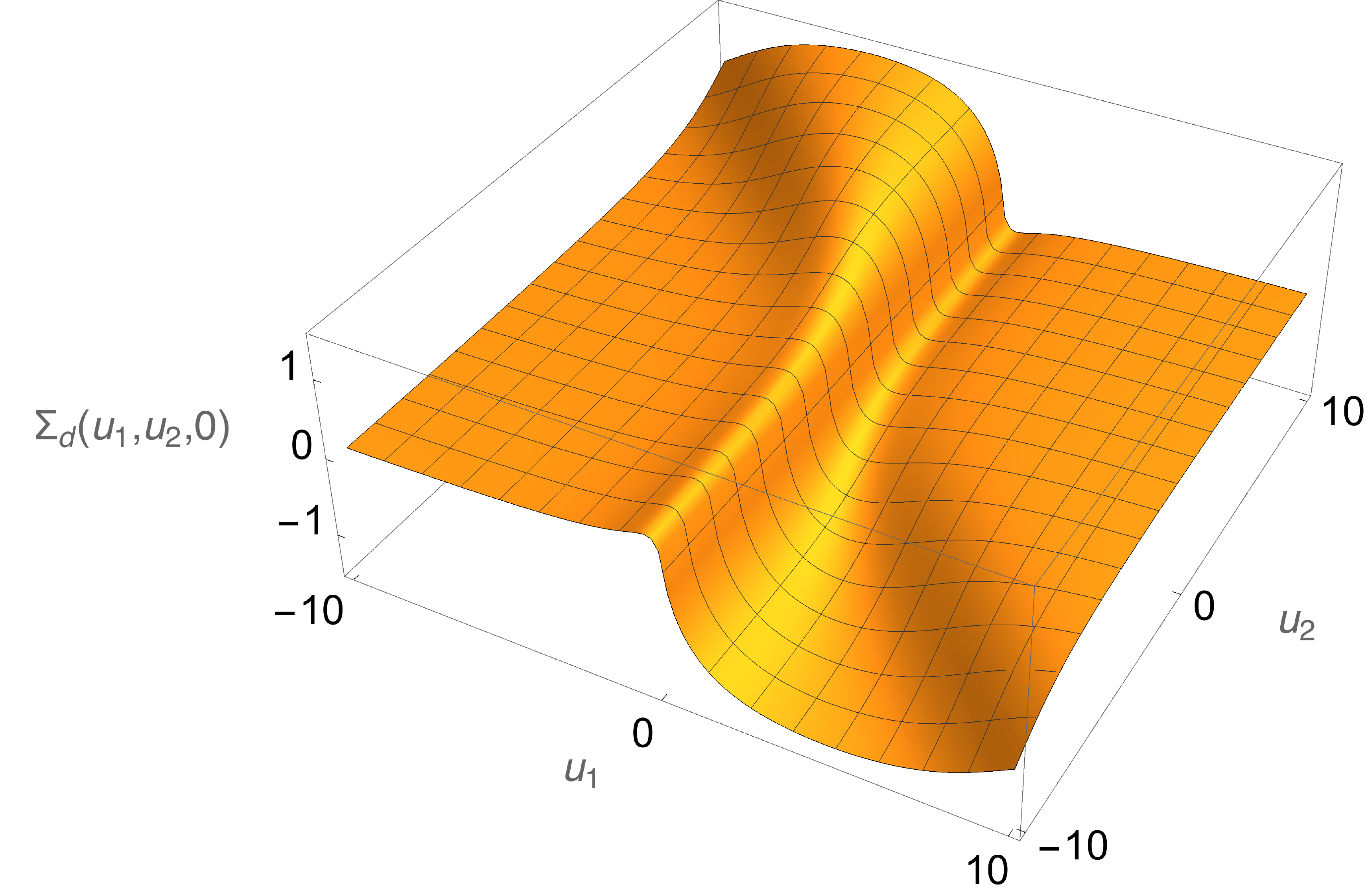}\hspace*{18mm}
    \includegraphics[width=57mm,angle=0]{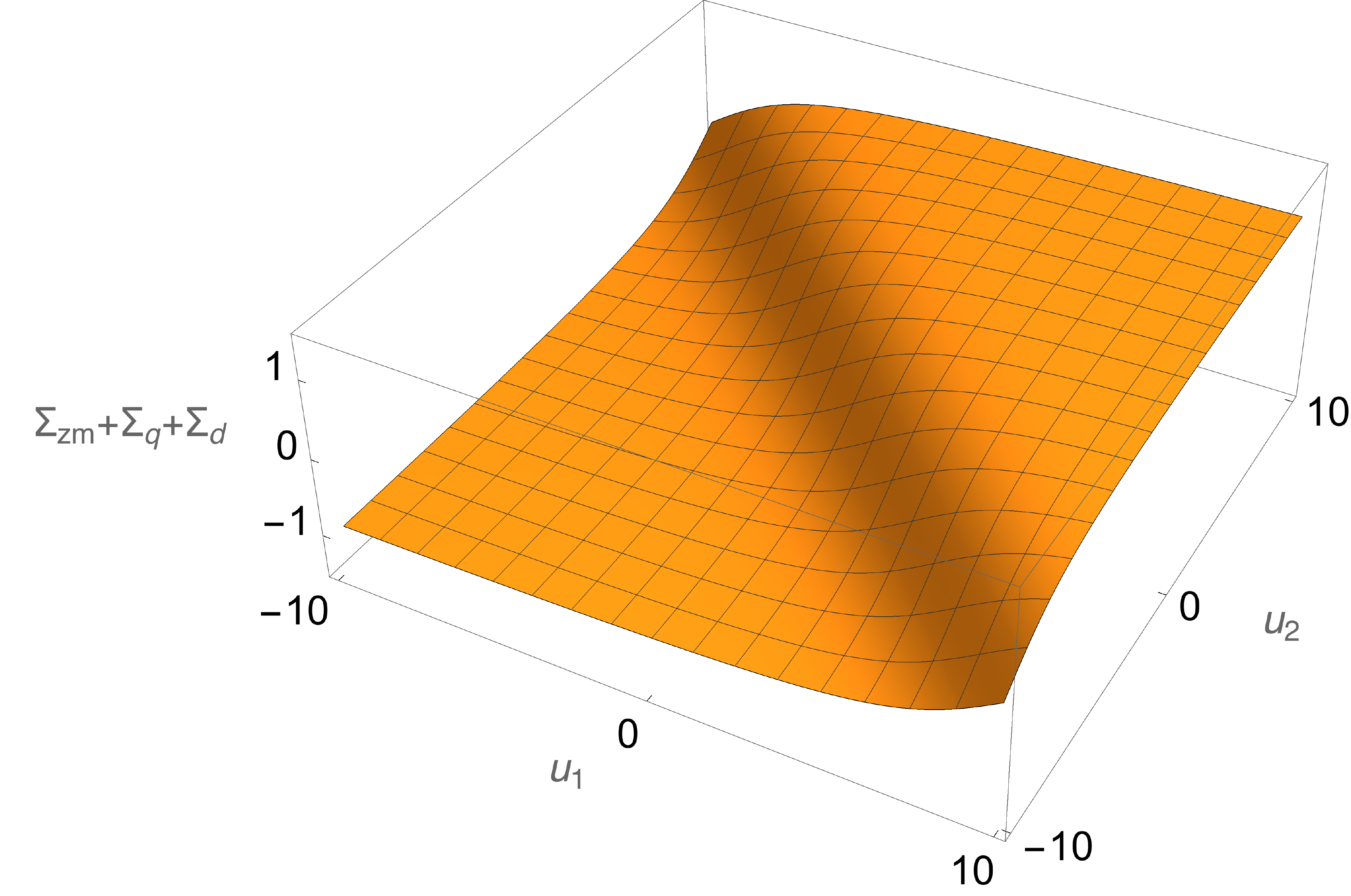}\vspace*{-7mm}
  \end{center}
  \caption{Contributions to the chiral condensate for two-flavor QCD
    at $\theta=0$.  The contributions of the quenched part and the
    zero-mode part of the spectral density increase exponentially with
    the volume when the product of the quark masses is negative (upper
    figures). However, their sum remains finite (central figure), and
    after adding the dynamical part we observe a discontinuity on the
    line $\hm_1 + \hm_2 =0$.}
  \label{fig8}
\end{figure}

For $N_f=2$ it is also possible to perform the sum over $\nu$ to
obtain the spectral density of the Dirac operator at fixed
$\theta$-angle. The nonzero-mode part reads
\begin{align}
  \rho_\nzm(\hx,\hm_1,\hm_2,\theta)
  &=\frac{|\hx|}{Z(\hm_1,\hm_2,\theta)}\int_{-\pi}^\pi\frac{d\phi}{2\pi}\Biggl\{
    \frac{J_1(2\hx\sin\frac\phi2)}{2\hx\sin\frac\phi2}Z(\hm_1,\hm_2,\theta-\phi)\notag\\
  &\quad+\frac{i\hx e^{i(\phi/2-\theta)}{(2\hm_1\hm_2-(\hm_1^2+\hm_2^2)e^{i(\theta-\phi)})}}
    {(\hx^2+\hm_1^2)(\hx^2+\hm_2^2)}J_1(2\hx\sin\tfrac\phi2)
    Z(\hm_1,\hm_2,\theta-\phi)   \label{eq:rho2nzm} \\
  &\quad -\frac{J_0(2\hx\sin\frac\phi2)}
    {(\hx^2+\hm_1^2)(\hx^2+\hm_2^2)}
    I_0\left(\sqrt{\hm_1^2+\hm_2^2+2\hm_1\hm_2\cos(\theta-\phi)}\right)
    \left(\hm_1\hm_2 e^{-i(\theta-\phi)}+\hx^2 e^{-i\phi}\right) \Biggr\}\,,\notag
\end{align}
where the first line corresponds to the quenched part and the rest to
the dynamical part. In Fig.~\ref{fig7} we show these two parts
separately as a function of $\hx$ and $\hm_1$ for $\hm_2 =10$ and
$\theta =0$. Again we observe exponentially large oscillations for
$\hm_1< 0$, this time only for negative quark mass because
$\theta = 0$.  The zero-mode contribution to the spectral density is
given by the principal-value integral
\begin{align}
  \label{eq:rho2zm}
  \rho_\zm(\hx,\hm_1,\hm_2,\theta) = -\delta(\hx) 
  \int_{-\pi}^\pi \frac{d\phi}{4\pi \sin^2\frac\phi2}\,
  \frac{Z(\hm_1,\hm_2,\theta-\phi)- Z(\hm_1,\hm_2,\theta)}{Z(\hm_1,\hm_2,\theta)}\,.
\end{align}
For large volume the integrand of \eqref{eq:rho2zm} is dominated by
$\phi=\tilde\theta$, with $\tilde\theta=\theta$ for $\hm_1\hm_2>0$ and
$\tilde\theta=\theta+\pi$ for $\hm_1\hm_2<0$, resulting in the
asymptotic contribution to the chiral condensate of the first quark
\begin{align}
  \Sigma_\zm(\hm_1,\hm_2,\theta) \sim -\frac 1{4\pi \hm_1 \sin^2\frac{\tilde\theta}2} 
  \frac {Z(\hm_1,\hm_2,\theta-\tilde\theta)}{Z(\hm_1,\hm_2,\theta)}
  \quad\text{(valid for $\tilde\theta$ away from 0 mod $2\pi$)}\,,
\end{align}
which grows exponentially with the volume since for $\tilde\theta \ne 0$
the free energy of the denominator is less than the free energy of the
numerator.  Since the sum of the zero-mode part and the quenched part
of the chiral condensate is finite for $ V\to \infty$, the quenched
part should also grow exponentially but with the opposite sign. It is
given by
\begin{align}
  \Sigma_q(\hm_1,\hm_2,\theta) = \int_{-\pi}^\pi d\phi\,
  \frac {1-2| \hm_1 \sin\frac \phi 2|K_1(2|\hm_1\sin\frac \phi 2|)}
  {4\pi \hm_1\sin^2\frac \phi 2 }
  \frac {Z_2(\hm_1,\hm_2,\theta-\phi) }{Z_2(\hm_1,\hm_2,\theta)}\,.
\end{align}
For large $V$ this simplifies to
\begin{align}
  \Sigma_q(\hm_1,\hm_2,\theta) \sim \frac 1{4\pi \hm_1 \sin^2\frac{\tilde\theta}2}
  \frac {Z_2(\hm_1, \hm_2,\theta-\tilde\theta)}{Z_2(\hm_1, \hm_2,\theta)}
  \quad\text{(valid for $\tilde\theta$ away from 0 mod $2\pi$)}\,,
\end{align}
which exactly cancels the exponentially increasing contribution from
the zero modes. One can show that this cancellation works to all
orders \cite{Kieburg:2017}. In Fig.~\ref{fig8} we show that both the
quenched and the zero-mode contributions to the chiral condensate grow
exponentially for $\hm_1\hm_2 <0$ (upper two figures), but that their
sum is finite (central figure).  The discontinuity shown in
Fig.~\ref{fig2} is recovered after adding the dynamical contribution
(lower figures).

\section{Conclusions}

In the $\epsilon$-domain of QCD we have obtained exact analytical
expressions for the eigenvalue density of the Dirac operator at fixed
$\theta \ne 0$ for both one and two flavors. These results made it
possible to explain how the different contributions to the spectral
density conspire to give a chiral condensate at fixed $\theta$ that
does not change sign when the quark mass (or one of the quark masses
for two flavors) crosses the imaginary axis, while the chiral
condensate at fixed topological charge does change sign.  From QCD at
nonzero density we have learnt that the discontinuity of the chiral
condensate may move to a different location when the spectral density
increases exponentially with the volume with oscillations on the order
of the inverse volume. This is indeed what happens when the product of
the quark masses becomes negative, but the situation is more subtle in
this case: the contribution of the ``quenched'' part of the spectral
density diverges in the thermodynamic limit at nonzero $\theta$, but
this divergence is canceled exactly by the contribution from the zero
modes.  We conclude that the zero modes are essential for the
continuity of the chiral condensate and that their contribution has to
be perfectly balanced against the contribution from the nonzero modes.
Lattice simulations at nonzero $\theta$-angle can only be trusted if
this is indeed the case.

This work was supported in part by DFG grant AK35/2-1 (MK), U.S.\ DOE
grant No.\ DE-FAG-88FR40388 and the Alexander von Humboldt Foundation
(JV), and DFG grant SFB/TRR-55 (TW).

\bibliography{sumnu}

\end{document}